\documentclass[article,longbibliography,twocolumn,reprint,amsmath,amssymb,aps,showpacs,superscriptaddress,prl]{revtex4-2}

\usepackage[utf8]{inputenc}
\usepackage{graphicx}
\usepackage{bm}
\usepackage{epsfig}
\usepackage{amssymb}
\usepackage{amsfonts}
\usepackage{braket}
\usepackage{color}
\usepackage{epstopdf}
\usepackage{hyperref}
\usepackage{float}
\restylefloat{table}
\usepackage{bibentry}
\usepackage{color}
\usepackage{multirow}
\usepackage[caption=false]{subfig}
\usepackage{makecell}

\DeclareUnicodeCharacter{0301}{\'{e}}

\newcommand{\bpm}{\begin{pmatrix}}
	\newcommand{\epm}{\end{pmatrix}}
\newcommand{\ba}{\begin{eqnarray}}
\newcommand{\ea}{\end{eqnarray}}
\newcommand{\bd}{\begin{displaymath}}
\renewcommand{\v}[1]{{\bf #1}}

\graphicspath{{figures/}}

\begin{document}
\title{Stable hump-like Hall effect and non-coplanar spin textures in SrRuO$_3$ ultrathin film}

\author{Byungmin Sohn}
\affiliation{Department of Physics and Astronomy, Seoul National University, Seoul 08826, Korea}
\affiliation{Center for Correlated Electron Systems, Institute for Basic Science, Seoul 08826, Korea}
\author{Bongju Kim}
\email[Electronic address:$~~$]{bongju@snu.ac.kr}
\affiliation{Department of Physics and Astronomy, Seoul National University, Seoul 08826, Korea}
\affiliation{Center for Correlated Electron Systems, Institute for Basic Science, Seoul 08826, Korea}
\author{Se Young Park}
\affiliation{Department of Physics and Astronomy, Seoul National University, Seoul 08826, Korea}
\affiliation{Center for Correlated Electron Systems, Institute for Basic Science, Seoul 08826, Korea}
\author{Hwan Young Choi}
\affiliation{Department of Physics, Yonsei University, Seoul 03722, Korea}
\author{Jae Young Moon}
\affiliation{Department of Physics, Yonsei University, Seoul 03722, Korea}
\author{Taeyang Choi}
\affiliation{Department of Physics, Chung-Ang University, Seoul 06974, Korea}
\author{Young Jai Choi}
\affiliation{Department of Physics, Yonsei University, Seoul 03722, Korea}
\author{Hua Zhou}
\affiliation{Advanced Photon Source, Argonne National Laboratory, Argonne, IL 60439, USA}
\author{Jun Woo Choi}
\affiliation{Center for Spintronics, Korea Institute of Science and Technology, Seoul 02792, Korea}
\author{Alessandro Bombardi}
\affiliation{Diamond Light Source Ltd., Harwell Science and Innovation Campus, Didcot, Oxfordshire, OX11 0DE, United Kingdom}
\affiliation{Department of Physics, University of Oxford, Parks Road, Oxford OX1 3PU, United Kingdom}
\author{Dan. G. Porter}
\affiliation{Diamond Light Source Ltd., Harwell Science and Innovation Campus, Didcot, Oxfordshire, OX11 0DE, United Kingdom}
\author{Seo Hyoung Chang}
\email[Electronic address:$~~$]{cshyoung@cau.ac.kr}
\affiliation{Department of Physics, Chung-Ang University, Seoul 06974, Korea}
\author{Jung Hoon Han}
\affiliation{Department of Physics, Sungkyunkwan University, Suwon 16419, Korea}
\author{Changyoung Kim}
\email[Electronic address:$~~$]{changyoung@snu.ac.kr}
\affiliation{Department of Physics and Astronomy, Seoul National University, Seoul 08826, Korea}
\affiliation{Center for Correlated Electron Systems, Institute for Basic Science, Seoul 08826, Korea}\date{\today}

\begin{abstract}
We observed a hump-like feature in Hall effects of SrRuO$_3$ ultrathin films, and systematically investigated it with controlling thicknesses, temperatures and magnetic fields. The hump-like feature is extremely stable, even surviving as a magnetic field is tilted by as much as 85$^\circ$. Based on the atomic-level structural analysis of a SrRuO$_3$ ultrathin film with a theoretical calculation, we reveal that atomic rumplings at the thin-film surface enhance Dzyaloshinskii-Moriya interaction, which can generate stable chiral spin textures and a hump-like Hall effect. Moreover, temperature dependent resonant X-ray measurements at Ru L-edge under a magnetic field showed that the intensity modulation of unexpected peaks was correlated with the hump region in the Hall effect. We verify that the two-dimensional property of ultrathin films generates stable non-coplanar spin textures having a magnetic order in a ferromagnetic oxide material.

\end{abstract}
\maketitle
\section{Introduction}

In magnetic materials, topological properties based on non-coplanar spin textures offer intriguing possibilities for exploring emergent properties and creating functionalities at the nanoscale. In ferromagnets, a transverse electric field can be induced without an external magnetic field owing to asymmetric spin scattering, known as anomalous Hall effect (AHE)~\cite{bruno06}. Recently, beyond AHE a non-vanishing anomalous conductance like a hump-like feature appears in the Hall effect~\cite{pyrochlore, nagaosa-review}. This emergent Hall effect, dubbed as topological Hall effect (THE), is found to be from real-space Berry curvature of non-coplanar spin or magnetic structures such as all-in-all-out spins in pyrochlore lattice~\cite{pyrochlore} or magnetic skyrmions~\cite{nagaosa-review, jiang-review, han-book, fert-review}. During the last decade, THE has been observed in materials with non-coplanar spin textures~\cite{MnSi_THE, MnSi_THE2, FeGe_THE, MnFeSi_THE}. 

A key ingredient for generating non-coplanar spin structures in the thin film systems~\cite{MnSi_thinfilm, FeGe_thinfilm} and heterostructures~\cite{BiSbTe, MgO} is well-known to be the Dzyaloshinskii-Moriya interaction (DMI) in the presence of an inversion symmetry breaking (ISB). The underlying design strategy for achieving the stability has been artificially breaking the inversion symmetry and boosting the spin-orbit coupling necessary for DMI by exploiting the heterostructure of thin film with a heavy-metal layer~\cite{fert2013skyrmions}. However, such heterostructures usually become inevitably quite complicated, which should be detrimental to construction of the theoretical models and designing highly scalable energy efficient devices.

SrRuO$_3$ (SRO), a well-known itinerant ferromagnet, was recently shown to be an excellent platform for studying THE~\cite{matsuno, ohuchi18} and engineering the stabilization of non-coplanar spin textures. For instance, a capping layer with Ir element grown on SRO films was able to enhance the spin-orbit coupling in SRO and to increase the DMI energy for THE. Moreover, researchers reported that in the ultrathin limit, SRO films exhibited the THE even without the additional heavy elements. Note that the phenomena can be described by topological effects or inhomogeneities in AHEs~\cite{Wu_2020, groenendijk2020berry}. For instance, a recent study on magnetic circular dichroism patterns verified the existence of chiral spin structure in SRO films~\cite{huang2020detection}. Despite these works, there is still a lack of deep understanding on how to stabilize the non-coplanar spin textures in the ultrathin limit.

\begin{figure*}[t!]
\includegraphics[width=\linewidth]{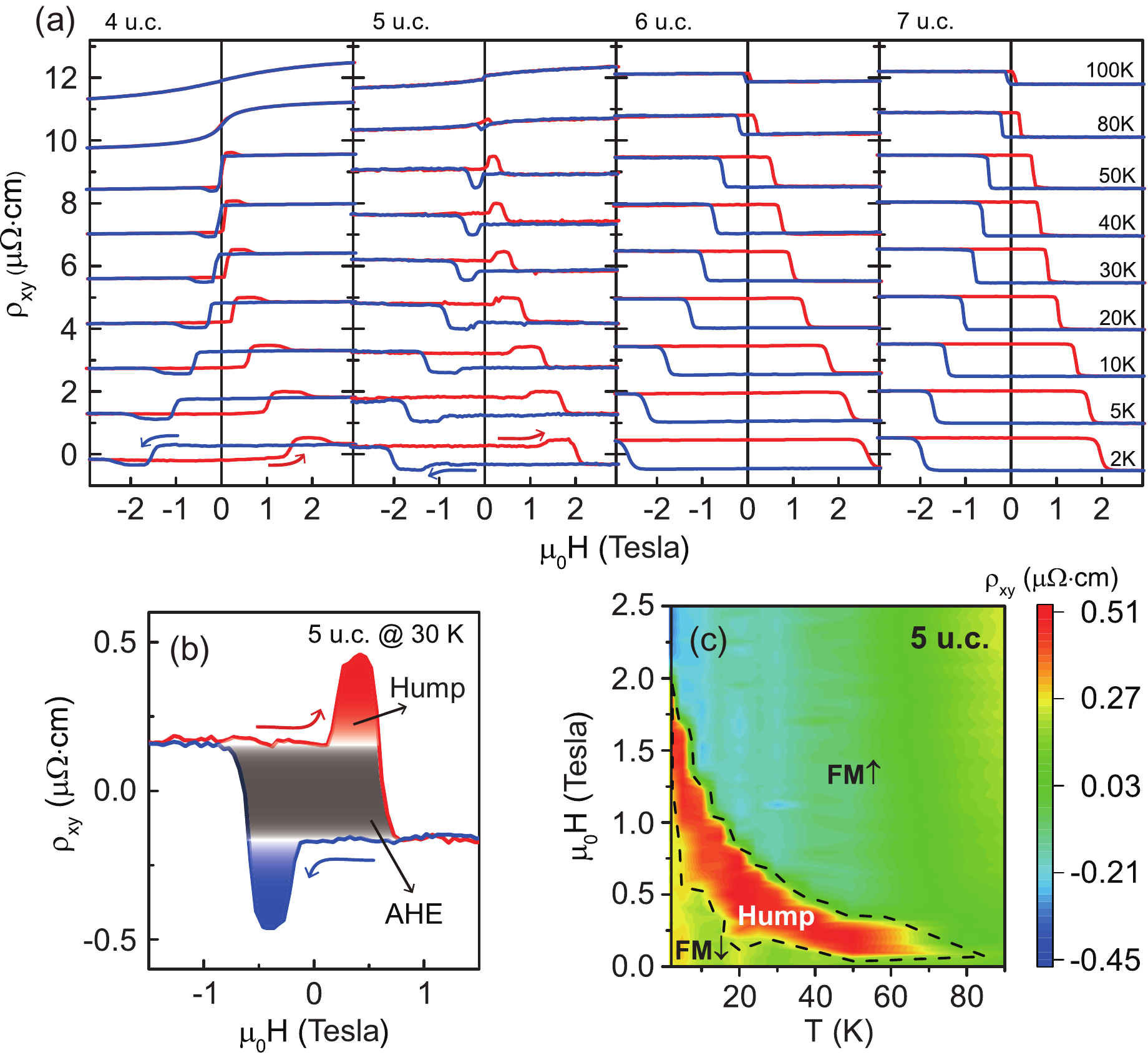}
\caption{Emergent hump-like structuers in Hall effect of SrRuO$_3$ (SRO) ultrathin films under the perpendicular magnetic field. (a) Hall effect measurement as a function of the perpendicular field $\mu_0 H$ at various temperatures for 4~-~7 unit cells (u.c.) SRO after subtraction of the ordinary Hall effect. The red (blue) arrow indicates the positive (negative) sweep direction of the magnetic field. Both anomalous Hall effect (AHE) and the hump-like features are observed on 4 and 5~u.c. SRO, while only AHE is observed for 6 and 7~u.c. SRO. (b) An enlarged plot of the Hall effect measured for 5~u.c. SRO at 30~K. The curves contain contributions from both AHE and the hump structure. (c) Phase diagram of 5~u.c. SRO in the $(T,\mu_0 H)$ plane. Color plots represent the measured (anomalous + hump) Hall resistivity values. Assignments of the phases are ferromagnetic (FM) and emergent hump-like structure (Hump) regions, respectively. The hump region is defined between $H_{c1}(\theta)$ and $H_{c2}(\theta)$ in Fig. 2(b). Arrows next to the 'FM' indicate the direction of magnetization.}
\label{fig:1}
\end{figure*}

\begin{figure*}[]
\includegraphics[width=\linewidth]{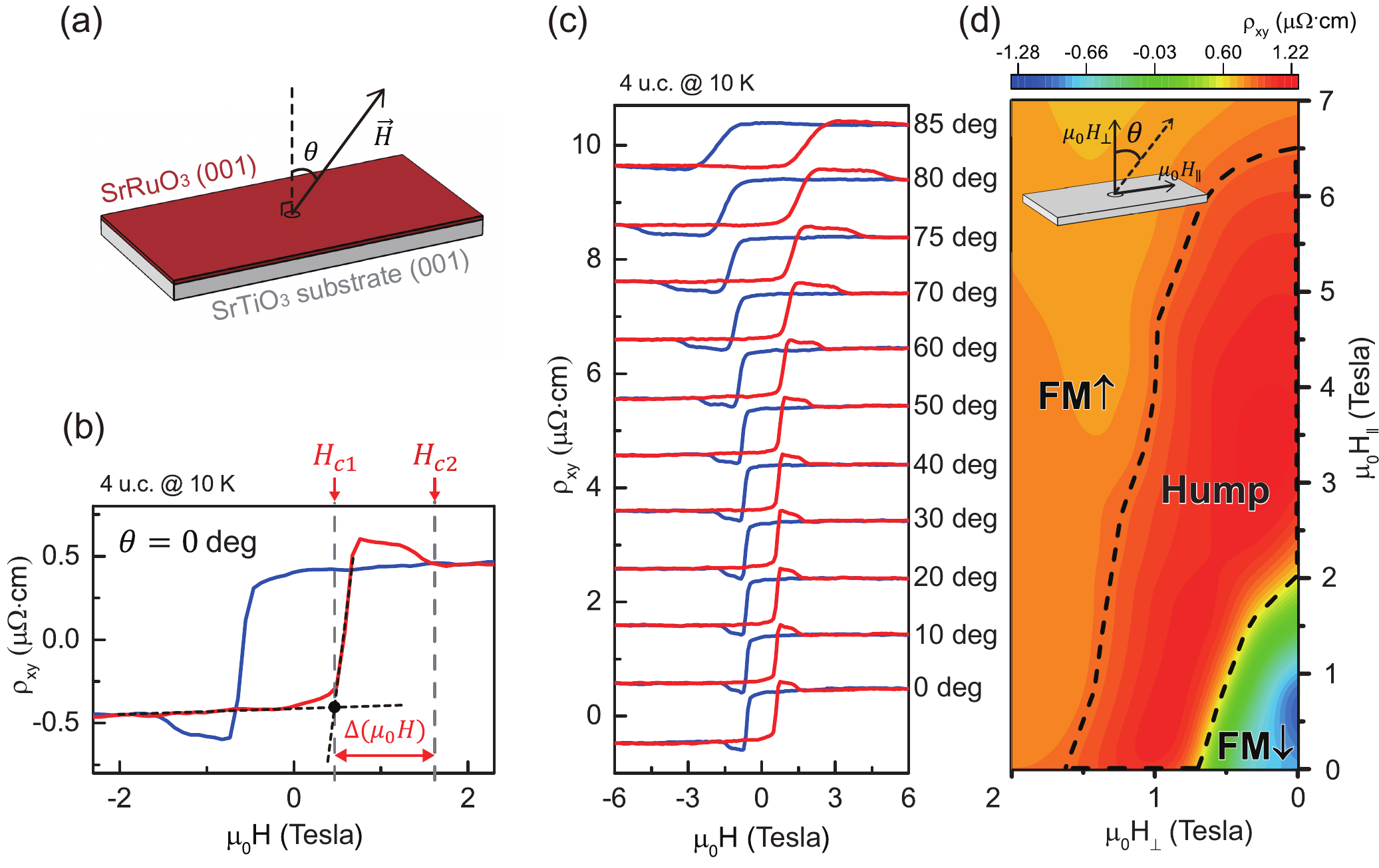}
\caption{Angle-dependent Hall measurements in 4~u.c. SRO. (a) Schematics of the geometry of angle-dependent Hall measurement. External magnetic field is tilted at an angle $\theta$ from the perpendicular direction. (b) Hall resistivity data when $\theta$ is zero. $H_{c1}$ is defined by the intersection of two extrapolations shown as black dotted lines. $H_{c2}$ is defined as the field value at which the Hall resistivity values under the positive and negative sweeps coincide. $\Delta (\mu_0 H )$ is defined as the difference, $\mu_0 (H_{c2} - H_{c1})$. (c) Hall resistivity at various angles of field inclination. (d) Phase diagram in the plane of parallel ($\mu_0 H_\parallel$) and perpendicular ($\mu_0 H_\perp$) components of the magnetic field. (Inset) A schematic of our experimental geometry with tilt angle, $\theta$.}
\label{fig:2}
\end{figure*}

\begin{figure*}[htbp]
\includegraphics[width=1\linewidth]{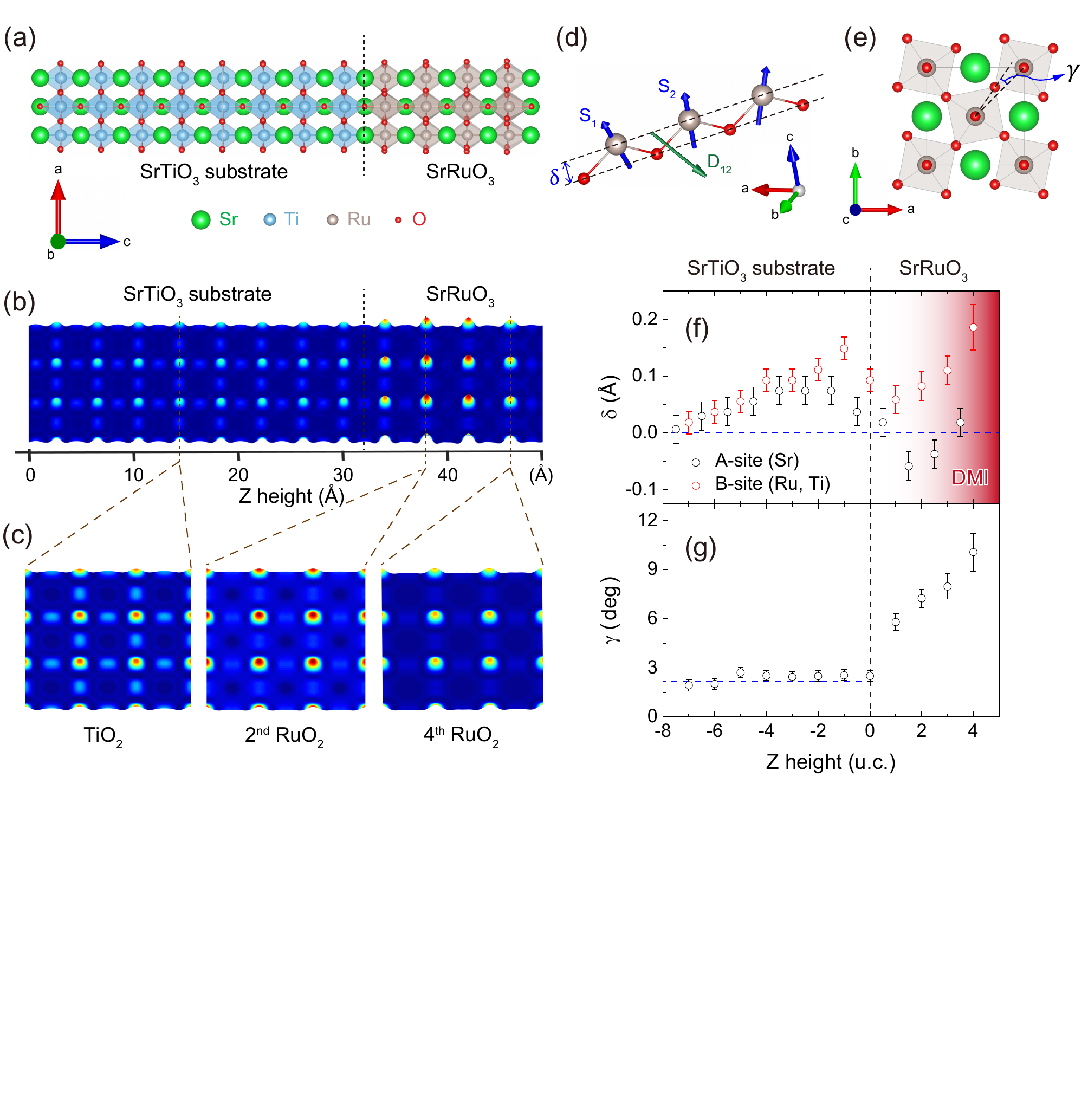}
\vspace{-6cm}
\caption{A source of Dzyaloshinski-Moriya interaction (DMI) in a SRO ultrathin film. (a) A crystal model of SRO on the SrTiO$_3$ (STO) substrate. A green (blue, gray, red) dot indicates Sr (Ti, Ru, O) atom. A blue (gray) translucent octahedron indicates a oxygen octahedron of Ti (Ru) atoms. SRO-STO interface is marked by black dotted line. (b) Electron density contour plot of STO substrate and 4~u.c. SRO in (100) plane at 30~K taken by coherent Bragg rod analysis (COBRA). (c) Electron density contour plots on B-site atoms for several Ti and Ru layers. (d) A schematic plot of the cation rumpling length, \textit{$\delta$}, measured from the center of the octahedron to the Ru atom, and the corresponding DMI vector (D$_{12}$). (e) SRO atomic structure within the $ab$-plane with the octahedral rotation angle, $\gamma$, between Ru atom and O atom. (f, g) Layer-dependent cation-oxygen rumpling, \textit{$\delta$}, and octahedral rotation angle, $\gamma$, of the STO substrate and the SRO thin film measured by COBRA. Blue dotted horizontal line indicates the average of \textit{$\delta$} and $\gamma$ for the STO substrate. The cation-oxygen rumpling near the film surface acts as a source of DMI.}
\label{fig:3}
\end{figure*}

\begin{figure*}[htbp]
\includegraphics[width=\linewidth]{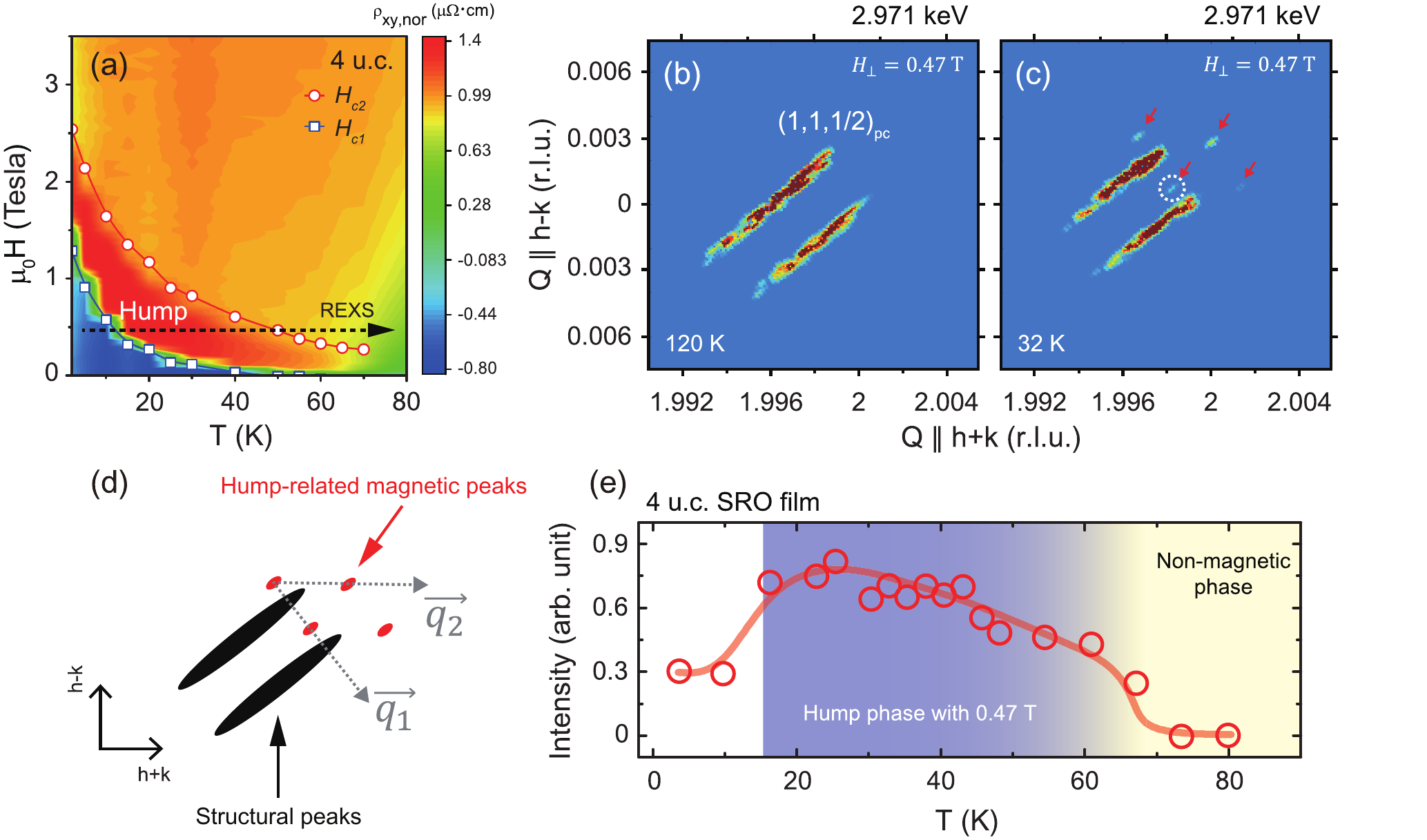}
\caption{Hump structure related magnetic order in a SRO ultrathin film. (a) Phase diagram of 4~u.c. SRO in the $(T,\mu_0 H)$ plane. Resonant elastic X-ray scattering (REXS) was performed as a function of temperature with $\mu_0 H$ = $0.47$~T. (b, c) Reciprocal space mapping (RSM) of REXS around SRO $(1,1,1/2)_{pc}$ reflection under the magnetic field of 0.47~T at 120~K and 32~K, where $pc$ represents for pseudocubic. Additional peaks are shown at 32~K (red arrows). (d) Schematic description of REXS features around SRO $(1,1,1/2)_{pc}$ reflection. Black stripy (red dotted) patterns are structural (hump-related magnetic) peaks. An incommensurate magnetic modulation vector $\vec{q_{1}}$ is alined along the direction of the step terrace, whereas $\vec{q_2}$ is tilted from the step terrace. (e) Temperature-dependent intensity of hump-related magnetic peaks (white dotted circle in Fig. 4(c)).}
\label{fig:4}
\end{figure*}

In this paper, we present a systematic approach for investigating a novel Hall effect in SRO ultrathin films on SrTiO$_3$ (STO) (001) substrates with controlling thicknesses from 7 unit cells (u.c.) to 4 u.c.. 
Clear hump-like features, which we attribute to a THE, were observed in the Hall resistivity of 4 and 5~u.c. SRO thin films. The hump-like features were found to be quite robust, surviving with the magnetic field tilting angle as large as 85$^\circ$. 
To elucidate microscopic origins of the hump-like feature, we analyzed atomic-level structure of a 4~u.c. SRO thin film by using surface X-ray  scattering combined with coherent Bragg rod analysis (COBRA) and density functional theory (DFT) calculations. 
A cation-oxygen rumpling into the out-of-plane direction can lead to sufficient DMI required for stabilizing non-coplanar spin structures. Furthermore, temperature-dependent resonant elastic X-ray scattering (REXS) studies on a 4~u.c. SRO thin film showed nontrivial magnetic order that concomitantly appears with the hump-like features. Based on a theoretical calculation and experimental observations, we suggest that the robust hump-like structure in Hall resistivity is induced by a non-coplanar spin textures, which can be stabilized by DMI and strong out-of-plane magnetic anisotropy.

\section{results}
\subsection{Hall effect measurements}

Figure 1(a) shows Hall resistivity measured at various temperatures for 4~-~7~u.c. SRO under the perpendicular magnetic field to the thin film. Hump-like structures are clearly observed in cases of 4 and 5~u.c. SRO, while only large hysteresis is observed in 6 and 7~u.c. SRO. 
The Hall resistivity in our ultrathin films can be decomposed by three terms, $\rho_{xy} = \rho_{\rm OHE} + \rho_{\rm AHE} + \rho_{\rm hump}$, namely the ordinary Hall effect (OHE), AHE, and hump-like structures, whereas only OHE and AHE are sufficient to explain Hall effect in 6 and~7 u.c. SRO. Note that the $\rho_{xy}$ contributed by OHE was subtracted in all the data shown in the paper. The hump-like structure starts to appear, together with AHE, below the Curie temperature $T_c$, which can be determined from the temperature-dependent longitudinal resistivity $\rho_{xx} (T)$ [See Appendix B for details].

As illustrated in Fig. 1(b), a clear hump-like feature is observed, and disappears near the coercive field where AHE undergoes an abrupt change. At the lowest measured temperature of 2~K, the hump-like structure persists over the 1~Tesla (T) range, demonstrating that the hump structure in SRO film is extremely robust compared to THE in other materials that host non-coplanar spin textures over much weaker fields~\cite{tokura10,FeGe_THE}. By using the Hall resistivity data, an overall temperature-field phase diagram for 5~u.c. SRO can be obtained, as shown in Fig. 1(c). The phase diagram exhibited the universal trend of the nontrivial topological phase in ferromagnetic films compared to those previous reports in the SrRuO$_3$/SrIrO$_3$ heterostructure~\cite{matsuno}, the thin-film chiral magnet Fe$_{1-x}$Co$_x$Si~\cite{tokura10}, and FeGe~\cite{FeGe_thinfilm}. Interestingly, on the other hand, the critical field for destructing the hump structure is much larger in our SRO ultrathin films.

An intriguing aspect of the hump-like structure in ultrathin SRO films is its exceptional stability or robustness under the tilting of the magnetic field. As shown schematically in Fig. 2(a), the Hall resistivity of 4~u.c. SRO was measured with the magnetic field tilted at an angle $\theta$ from the perpendicular direction at 10~K. The angle-dependent Hall data are plotted in Fig. 2(c), using the total magnetic field ($\mu_0 H$) as the variable for the plot. At each angle of inclination one can define two critical fields $H_{c1}(\theta)$ and $H_{c2}(\theta)$, associated with the onset and disappearance of the hump, respectively (Fig. 2(b)). The stable region of hump-like structure is then defined as $\Delta (\mu_0 H ) \equiv \mu_0 (H_{c2} - H_{c1})$ for a given angle of inclination. $H_{c1}(\theta)$, $H_{c2}(\theta)$, and $\Delta (\mu_0 H )$ all increased continuously as the tilt angle increases.
Figure 2(d) represents a phase diagram in the plane of ($\mu_0 H_\parallel$, $\mu_0 H_\perp$), deduced from the Hall effect measurement data. The two components refer to the in-plane ($\mu_0 H_\parallel$) and out-of-plane ($\mu_0 H_\perp$) magnetic field, respectively. There exists a marked increase in the magnetic field window of stability for the hump structure as the tilt angle $\theta = \tan^{-1} (H_\parallel / H_\perp )$ grows.

Earlier experiment on a EuO film found the hump-like structure disappears at $\theta$ slightly larger than $10^\circ$ inclination of a external magnetic field from the normal~\cite{ohuchi}. Similar field-tilt measurement on the FeGe found $\theta_c \approx 12^\circ$ to be the critical angle of inclination before the cycloidal phase replaced the skyrmion phase~\cite{che}. Compared to similar field-tilt experiments in EuO~\cite{ohuchi}, FeGe~\cite{che}, GaV$_4$S$_8$~\cite{loidl15} and GaV$_4$Se$_8$~\cite{loidl17}, our ultrathin SRO shows an exceptionally wide phase diagram region of the hump-like structure under the field tilting, $e.g.,$ 85$^\circ$. 

\subsection{Surface cation rumpling responsible for the humplike features}

If the DMI is responsible for the hump-like feature, ISB may appear in the atomic structure. Hence, atomically resolved surface X-ray diffraction measurements combined with COBRA method render some key insights into the origin of the robust hump-like feature~\cite{hua10, shin17, fister14}. Figure 3(a) schematically illustrates the crystal structure of the 4~u.c. SRO thin film grown on STO (001) substrate. The COBRA method, being an effective phase-retrieval surface X-ray technique for ultrathin films, is capable of providing the three-dimensional electron density profile and accurate atomic positions. The overall two-dimensional electron density profiles of the 4~u.c. SRO and the STO layers are seen in the (100) plane for the Ru, Ti, and O atoms (Fig. 3(b)). Electron density profiles in some of the individual Ti-O$_2$ and Ru-O$_2$ planes are shown in Fig. 3(c) in greater detail.

The SRO ultrathin film on the STO substrate maintains the tetragonal structure $(a^0a^0c^-)$~\cite{SHChang} [See Appendix C for the low energy electron diffraction (LEED) data], meaning that the octahedral tilt angle along in-plane crystallographic axes is zero, while the rotation about the $c$-axis (by angle $\gamma$) is allowed. The intriguing aspect of the COBRA findings is the cation(Ru)-oxygen rumpling, measured by the displacement, $\delta$, of the Ru atom out of the RuO$_2$ plane (Figs. 3(d) and 3(e)). Both these quantities have been measured for the individual SRO and STO layer (Figs. 3(f) and 3(g)). A significant ionic displacement of the SRO ultrathin film observed by COBRA accounts for the ISB and the consequent appearance of DMI in the SRO thin film, which is consequently expected to produce non-coplanar spin structures. 


In order to quantitatively estimate the DMI energy, we performed first-principles calculation using a supercell with average value of the octahedral rotation angles and rumpling parameters of $7.8^{\circ}$ and 0.11 {\AA}, respectively, measured by COBRA [See Appendix A for details]. As displayed in Fig. 3(d), the direction of the DMI vector, ${\bf D}_{12}$, is perpendicular to the Ru($\vec{S_1}$)-O-Ru($\vec{S_2}$) plane and its strength is also proportional to the cation(Ru)-oxygen rumpling length~\cite{cheong17}, which produces antisymmetric exchange interaction, $H_{ij}=D_{ij} \cdot (S_i \times S_j)$. We obtained the DMI energy of $2.1$ meV, which is comparable to interfacial DMI energy for various metallic ferromagnet/heavy metal bilayer systems ($\sim1$ meV)~\cite{Nembach15, Chen13}. Thus, the DMI energy in SRO ultrathin films is large enough to induce non-coplanar magnetic structure in good agreement with the estimated DMI energy in SrRuO$_3$/SrIrO$_3$ heterostructures~\cite{matsuno}. Angle-dependent Hall resistance measured at 10 K at the magnetic field of 1.5 T and 6 T verified that SRO ultrathin films have perfect out-of-plane magnetic anisotropy [See Appendix D for details]. Two-dimensionality of ultrathin SRO film can induce the strong magnetic anisotropy along the normal direction, which contributes to the robust hump-like feature.


\subsection {Nontrivial magnetic order related to the humplike features}

It has been demonstrated that hump structures in Hall measurement could be originated from the non-coplanar spin order such as skyrmion lattice~\cite{MnSi_THE,zang-theory}. Thus, hidden magnetic orders might emerge with hump-like features in a 4~u.c. SRO ultrathin film. REXS is a powerful experimental method to directly observe a magnetic order in reciprocal space~\cite{zhang17}. We measured temperature-dependent REXS with a permanent magnet which induces magnetic field, $\mu_0 H$ $\sim$ 0.47~T, perpendicular to the thin film. The REXS measurement is performed at Ru L$_2$ edge ($h\nu = 2.971$~keV). Figure 4(a) represents the experimental range of REXS in the temperature-field phase diagram of Hall measurement for the 4~u.c. SRO film. 

Figures 4(b) and 4(c) show reciprocal space mapping (RSM) results of REXS measurement under 0.47~T at 120 and 32~K, respectively. At $(1,1,1/2)_{pc}$ ({\it pc} denotes for pseudocubic), we observed very weak and nearly temperature-independent diffraction patterns, which can originate from the crystal truncation rods or cation displacement. As shown in Fig. 4(b), only stripy patterns are observed at 120~K. The separation between the stripes in the reciprocal space is 0.0028~$\AA^{-1}$, and can be converted into the distance of $\sim220$~nm in real space, which is quite consistent with the width of step terraces measured by our atomic force microscopy [See Appendix B for details]. 

As decreasing the temperature to 32~K, the SRO film entered into hump phase (Fig. 4(a)). Then, extra peaks around the stripes appear in the RSM as pointed by red arrows in Fig. 4(c). Figure 4(d) illustrates that black and red peaks resulted from the step terraces and emergent magnetic order at low temperature, respectively. A magnetic vector, $\vec{q_{1}}$, is aligned parallel to the step-terrace direction, whereas $\vec{q_{2}}$ is tilted from the step-terrace direction. 

To confirm the correlation between hump phase and the emergent magnetic peaks, we plotted the intensity of emergent magnetic peaks as a function of temperature (Fig. 4(e)). The phase cross-over temperatures from hump to ferromagnetic and from ferromagnetic to non-magnetic phases were approximately 15~K and 60~K, respectively, with $\mu_0 H$ $\sim$ 0.47~T. At both of the cross-over temperatures, we clearly observed the abrupt change of intensity of the emergent magnetic peaks. Considering the strong correlation between the hump phase and the intensity of emergent peaks, we propose that the emergent magnetic order is responsible for the hump-like feature.

From the experimental results from transport, REXS, and COBRA with a theoretical calculation, we can conjecture the possible magnetic structure arising with the hump in Hall data. The magnetic structure should be periodic in a scale of step terrace based on the REXS result. COBRA and first-principle calculation present the source of DMI which can induce non-coplanar magnetic structures with large perpendicualr magnetic anisotropy~\cite{Thiaville12}. Recently, we observed a magnetic dichroism patterns in a SRO ultrathin film by using X-ray dichroism~\cite{huang2020detection}. Considering previous reports on the hump feature in the Hall measurement of SRO systems~\cite{matsuno, ohuchi18, Fengyuan19, Sohn20} and other oxide thin films~\cite{Bibes19, Yao19}, we claim that the non-coplanar spin order appears in SRO ultrathin films and induces hump-like structure in Hall measurement.

\section{Conclusion}
In summary, we performed Hall effect measurements in 4, 5, 6 and 7~u.c. SRO thin films grown on STO (001) substrate and observed hump-like structures in Hall measurements of 4 and 5 u.c. SRO thin films. 
Angle-dependent Hall effect measurement shows that the hump is extremely robust against the magnetic field angle, surviving even in the high angle as large as 85$^\circ$. 
To verify the origin of hump structures, layer-by-layer atomic structure of 4 u.c. SRO thin film is studied by COBRA which identifies Ru-O rumpling into the out-of-plane direction as the origin of DMI, which possibly stabilizes non-coplanar spin texture. 
Furthermore, we performed REXS under the magnetic field and observed non-trivial magnetic order emergent with the hump-like feature.
Based on the experimental observations and theoretical analysis with recent studies~\cite{huang2020detection, Sohn20}, we believe that the hump-like feature in Hall measurement is attributed to the non-coplanar spin order, and can be interpreted as THE.



\section{acknowledgments}

This work is supported by IBS-R009-G2 through the IBS Center for Correlated Electron Systems. 
T.C. and S.H.C. were supported by Basic Science Research Program through NRF (2019K1A3A7A09033393 and 2020R1A5A1016518). 
The use of the Advanced Photon Source at the Argonne National Laboratory was supported by the US DOE under Contract No. DE-AC02-06CH11357. The work at beamline I16 of Diamond Light Source was performed under Proposals MM22181. PPMS measurements were supported by the National Center for Inter-University Research Facilities (NCIRF) at Seoul National University in Korea. The work at Yonsei University was supported by the National Research Foundation of Korea (NRF) Grant (NRF-2017R1A5A1014862 (SRC program: vdWMRC center) and NRF-2019R1A2C2002601). J.W.C. acknowledges KIST Institutional Program.

\begin{center}
{\bf Appendix A: Methods}
\end{center}

SrRuO$_3$ (SRO) ultrathin films were grown on top of SrTiO$_3$ (STO) (001) substrate by pulsed laser deposition (PLD) technique. The STO substrate was prepared by deionized (DI) water etching and in-situ pre-annealing at $1070\,^{\circ}{\rm C}$ for 30 minutes with oxygen partial pressure (PO$_2$), $5\times$10$^{-6}$ Torr. Epitaxial SRO thin film was deposited with PO$_2$ = 100 mTorr at $700\,^{\circ}{\rm C}$. A KrF Excimer laser (wavelength = 248~nm) was delivered on stoichiometric SRO target with 1-2~J/cm$^2$ and repetition rate of 2~Hz. A 60~nm-thick Au top electrode with a Hall bar geometry was prepared on top of the SRO thin film by e-beam evaporator. Electric transport measurement was carried out by the Physical Property Measurement System (PPMS), Quantum Design Inc..

Resonant elastic X-ray scattering (REXS) experiment was performed at the beam line I16 of the Diamond Light Source (Didcot, UK). REXS was measured at the Ru L$_{2}$ absorption edge with linear polarization. A 6-circle diffractometer allowed us to have a grazing incidence diffraction geometry with an incident angle of 1.5$^\circ$ around the forbidden reflection $(1,1,1/2)_{pc}$. The X-ray path length at this energy and angle is roughly 30~nm and in this geometry the horizontal X-ray polarization is $\sim$60 degrees from the scattering plane. Air scattering was reduced with use of a helium-filled bag and an ultra-high gain, vacuum-enclosed Pilatus3-100K area detector. A closed-cycle cryocooler was used to control temperature. SRO thin films were mounted directly to a copper plate using a conductive silver paint. A permanent magnet was mounted directly below the plate, with a measured field on the sample $H = 0.47$~T, providing a steady magnetic field in order to measure the temperature dependence of the topological Hall effect in this system.

COBRA measurements were performed at Sector 33-ID-D of the Advanced Photon Source at Argonne National Laboratory. The samples were characterized at 30~K using a closed-cycle He cryostat (Displex). The crystal truncation rods (CTR) were measured with a Pilatus 100K area detector at an X-ray energy at 21.0~keV. The background were removed using the area detector. The CTR measurements were taken at 30~K in order to elucidate the detailed atomic structure of ultrathin SRO film. All the specular and off-specular CTR measurements were quantified using the COBRA method. The complex structure factors from measured CTR intensities were able to determine the electron density distribution with sub-Angstrom resolution using a Fourier transformation and iterative procedure~\cite{hua10}. Bulk SRO and thick SRO films grown on STO substrates at the room temperature or lower exhibit orthorhombic (a$^-$a$^-$c$^+$) or orthorhombic-like (monoclinic, a$^+$b$^-$c$^-$) structure, respectively, and the lack of cation-oxygen rumpling~\cite{koster12, jones08, gao16}.

First-principles density functional theory (DFT) calculations were performed with the generalized gradient approximation plus $U$ (GGA+$U$) method using the Vienna {\it ab-initio} simulation package (VASP)~\cite{PRB-1996-Kre,PRB-1999-Kre}. The projector augmented wave method~\cite{PRB-1994-Blo} was used with pseudopotentials containing 6 valence electrons for O ($2s^{2}2p^{4}$), 12 for Ti ($3s^{2}3p^{6}3d^{2}4s^{2}$), 10 for Sr ($4s^{2}4p^{6}5s^{2}$), and 14 for Ru ($4p^{6}4d^{7}5s^{1}$). The Perdew-Becke-Erzenhof parametrization~\cite{PRL-1996-Perdew} for the exchange-correlation functional and the rotationally invariant form of the on-site Coulomb interaction~\cite{PRB-1995-Lie} were used with $U = 1.4$ and $J=0.4$ eV for the Ru-$d$ orbitals~\cite{Rondinelli2008,Kim2014} and $U = 4$ and $J=0.68$~eV for Ti-$d$ orbitals~\cite{EPL-2008-Zhong,EPL-2011-Okatov,Science-2011-Jang}. Our choice of $U$ and $J$ values correctly reproduced the ferromagnetic metallic ground state of the bulk SRO with magnetization of 1 $\mu_{B}$/Ru aligned in plane consistent with experimental observations~\cite{Kanbayasi1978,Bushmeleva2006}. The parameter also reproduce the ferromagnetic metallic ground state of thin-film geometry of SRO with the magnetic anisotropy favoring the out-of-plane direction. We used the energy cutoff of 500 eV and $k$-point sampling on a $8\times 8\times 1$ grid with a $\sqrt{2}\times\sqrt{2}$ in-plane unit cell to include $a^{0}a^{0}c^{-}$ octahedral rotation of SRO/STO heterostructure. The detail of the methods used to calculate the magnetic anisotropy energy and DMI vector is presented in Appendix E.

\begin{figure}[htbp]
	\includegraphics[width=0.48\textwidth]{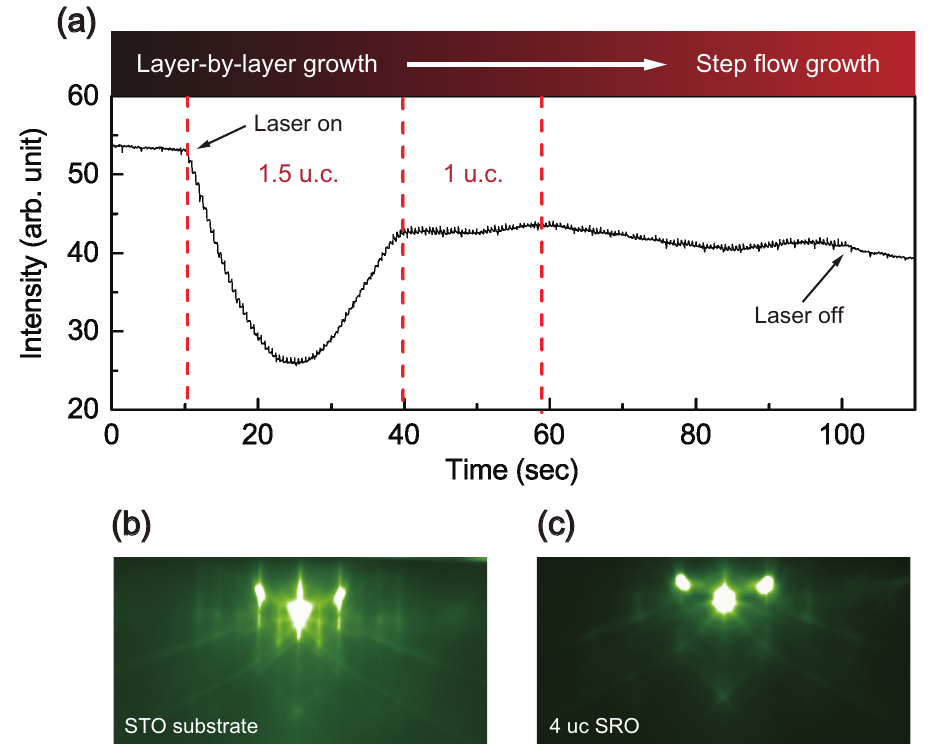}
	\caption{{\bf {\it In-situ} Reflection High-energy Electron Diffraction (RHEED) data of 4 u.c. SRO thin film.} (a) {\it In-situ} RHEED intensity versus time during SRO thin film growth. A clear growth mode transition from layer-by-layer growth to step flow growth is shown after 1.5 u.c. thickness of SRO is grown. The number of laser pulses for the growth of 1 u.c. is 38 pulses, from which we can calculate the repetition rate as 2 Hz and a growth rate of 0.103 {\AA}/sec. (b) A RHEED pattern of TiO$_2$-terminated STO substrate in UHV at 670 K. (c) The RHEED pattern of 4 u.c. SRO thin film in $10^{-8}$ mTorr at room temperature.}
	\label{RHEED}
\end{figure}

\begin{figure}[htbp]
	\centering
	\includegraphics[width=0.48\textwidth]{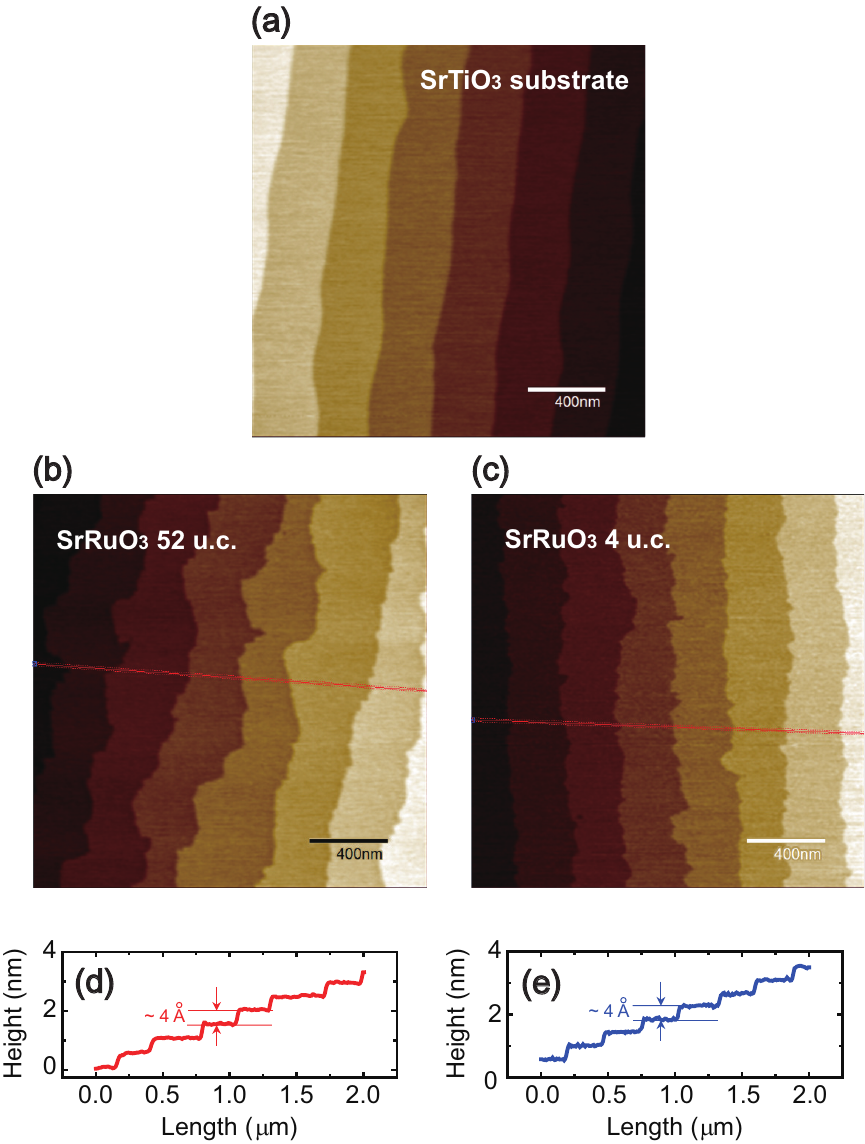}
	\caption{{\bf Atomic Force Microscopy (AFM) image of SRO thin film.} (a) AFM image of STO substrate. (b) AFM image of 52 u.c. SRO. (c) AFM image of 4 u.c. SRO. (d) AFM line profile of 52 u.c. SRO. (e) AFM line profile of 4 u.c. SRO. Clear step terrace structures are observed for all of the surfaces.}
	\label{AFM}
\end{figure}

\begin{figure}[htbp]
	\centering
	\includegraphics[width=0.48\textwidth]{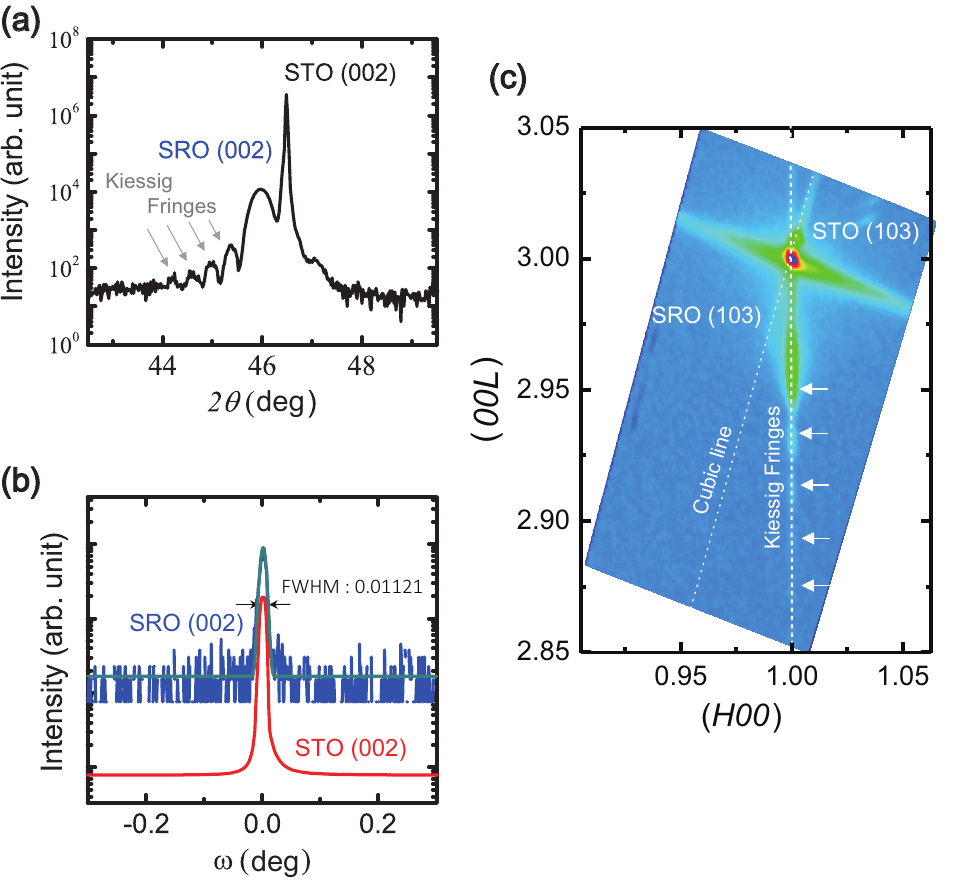}
	\caption{{\bf X-ray Diffraction (XRD) measurements of 50 u.c. SRO thin film.} (a) XRD $\theta - 2 \theta$ scan for SRO thin film. Kiessig fringes from the SRO thin film are clearly observed. (b) XRD rocking curve measurement of STO substrate and SRO thin film. The FWHM of SRO thin film is 0.01121$^\circ$, which shows the high crystallinity of SRO thin film. (c) Reciprocal space mapping of the SRO (103) and STO (103) peaks. Both peaks are aligned on the same H value, which shows that SRO thin film is fully strained by the STO substrate.}
	\label{XRD}
\end{figure}

\begin{figure}[htbp]
	\centering
	\includegraphics[width=0.47\textwidth]{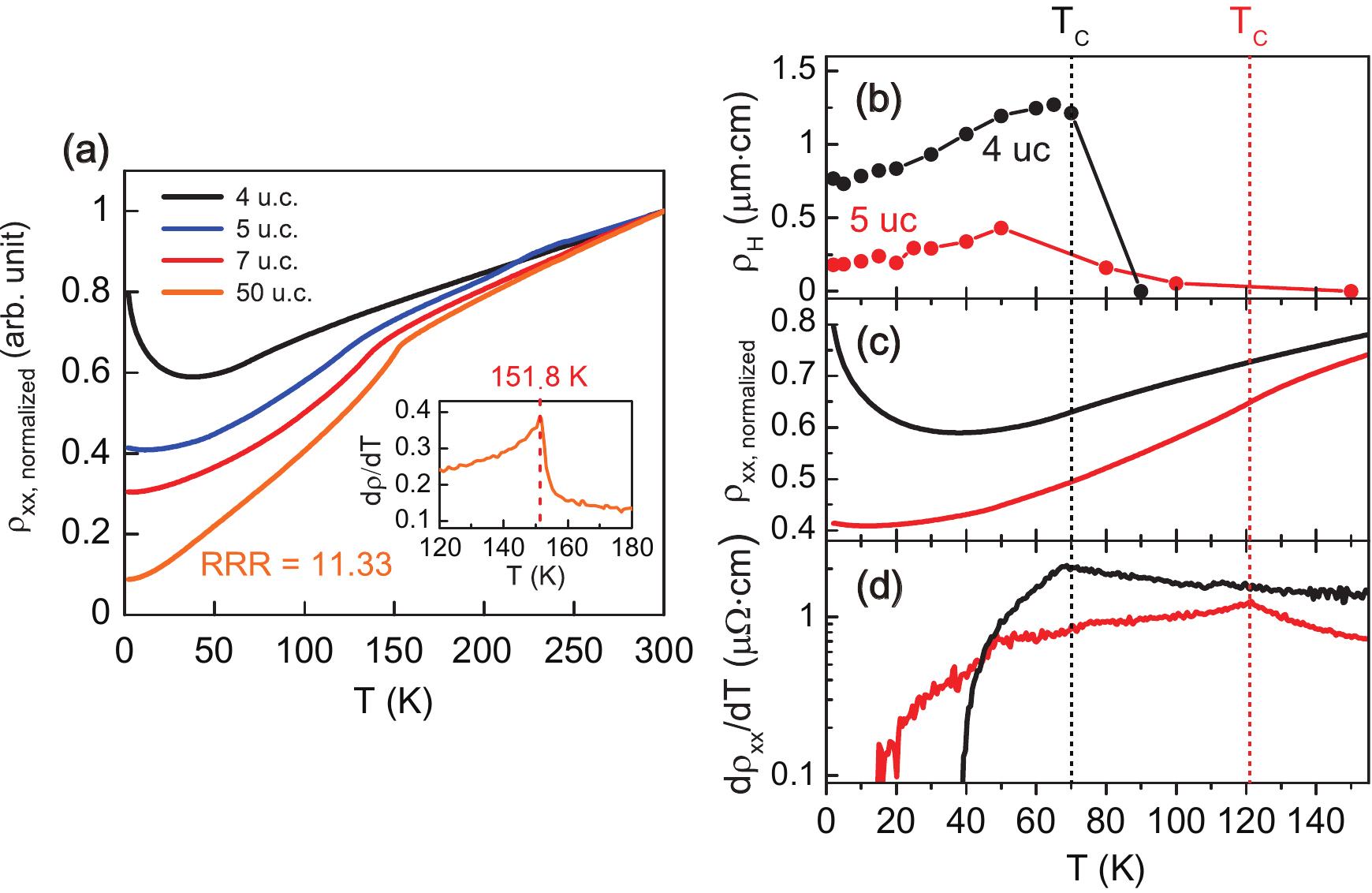}
	\caption{{\bf Resistivity of SRO thin film with various thicknesses.} (a) Normalized resistivity of 4, 5, 7 and 50~u.c. SRO thin films. The 50~u.c. SRO thin film exhibits a high residual resistivity ratio (RRR) value of 11.33. (Inset) The derivative of 50~u.c. SRO thin film resistivity. Curie temperature, $T_C$, is determined from the anomaly of the derivative curve. (b) Hump-like Hall resistivity of 4 and 5~u.c. SRO. $\rho_{H}$ is defined as the difference from the maximum of $\rho_{xy}$ to $\rho_{AHE}$. (c) Normalized resistivity of 4 and 5~u.c. SRO. (d) The derivative of 4 and 5~u.c. SRO resistivity. Curie temperature, $T_C$ is determined from the anomaly of the derivative curve.}
	\label{Res}
\end{figure}

\begin{center}
	{\bf Appendix B: Film characterization}
\end{center}

We performed {\it in-situ} reflection high-energy electron diffraction (RHEED), atomic force microscopy (AFM), X-ray diffraction (XRD), and resistivity to ensure the high quality of our SRO thin films. We have monitored the RHEED intensity versus time along the [100]$_c$ azimuth ($c$ indicates the cubic index) during the SRO growth, and clearly observed the transition in the growth condition from the layer-by-layer growth to the step flow growth as previous studies have reported~\cite{step-flow} (Fig.~\ref{RHEED}). AFM images of the SrTiO$_3$ (STO) substrate and SRO thin film surfaces are shown in Fig.~\ref{AFM}. A clear step terrace structure is observed for the STO substrate, 52~unit-cell (u.c.) SRO and the 4~u.c. SRO with the terrace height of 4~${\AA}$.

Figure~\ref{XRD} shows the XRD measurement for the 50~u.c. SRO thin film. In Fig.~\ref{XRD}(a), the range of 2~$\theta$ is from 42.5$^{\circ}$ to 49.5$^{\circ}$ to show the (002) peak of the STO substrate and SRO. Clear Kiessig fringes which appear due to an interference between the film surface and the film/substrate interface were observed. Figure~\ref{XRD}(b) shows the rocking curve measurement. The full width at half maximum (FWHM) of SRO (002) peak is given by 0.01121$^{\circ}$, which proves that the crystallinity of our SRO thin film is quite high. Figure~\ref{XRD}(c) shows the reciprocal space mapping near the symmetric (103) reflection of SRO. SRO and STO substrates are aligned at a single $H$ value, which shows that the SRO film is fully strained by the STO substrate.

We show in Fig.~\ref{Res}(a) the resistivity from 4, 5, 7 and 50~u.c. SRO thin films. A high value of residual resistivity ratio (RRR = 11.33) is obtained for the 50~u.c. SRO. We determined the Curie temperature to be $T_c$~=~151.8~K for the 50~u.c. SRO from the anomaly in the derivative curve. We define and plot $\rho_{H}$ of 4 and 5~u.c. SRO films in Fig.~\ref{Res}(b). The hump-like Hall effect is observed below the Curie temperature of each SRO film. Figure~\ref{Res}(d) shows the derivative of 4 and 5~u.c. SRO resistivity (Fig.~\ref{Res}(c)). The Curie temperature of each film is defined by the anomaly of the derivative curve.

These results show that our growth condition for SRO thin film by PLD method is well-optimized and the resulting SRO films are of very high quality.

\begin{center}
{\bf Appendix C: Structure analysis of SRO thin film by \textit{in-situ} LEED and RHEED}
\end{center}

\begin{figure}[htbp]
	\centering
	\includegraphics[width=0.48\textwidth]{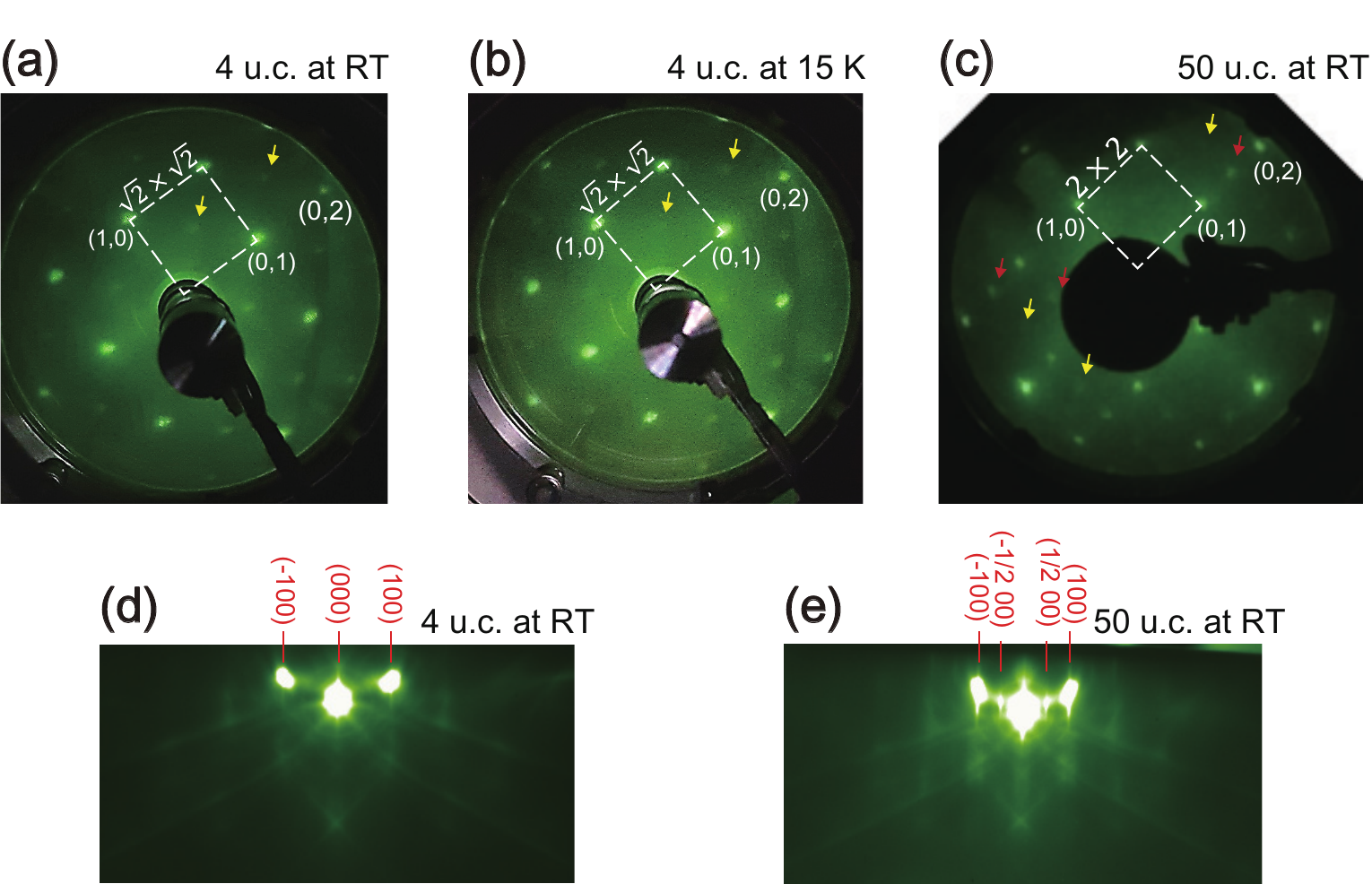}
	\caption{{\bf \textit{In-situ} low-energy electron diffraction (LEED) and reflection high-energy electron diffraction (RHEED) patterns of SRO thin film.} (a), (b) LEED patterns of 4 u.c. SRO at room temperature and 15 K with the electron beam energy of 142 eV. Yellow arrows indicate $\sqrt{2} \times \sqrt{2}$ peaks which are induced by tetragonal symmetry ($a^{0}a^{0}c^{-}$). (c) LEED pattern of 50 u.c. SRO at room temperature with the electron beam energy of 150 eV. Red arrows indicate $2 \times 2$ peaks which are induced by orthorhombic symmetry ($a^{-}a^{-}c^{+}$). (d) RHEED image of 4 u.c. SRO at room temperature. (e) RHEED image of 50 u.c. SRO at room temperature.}
	\label{LEED}
\end{figure}

After the SRO thin films were synthesized by PLD, films were characterized \textit{in-situ} by low-energy electron diffraction (LEED). LEED patterns of 4 and 50~u.c. SRO thin films are shown in Fig.~\ref{LEED}(a), (b), and (c). Figure~\ref{LEED}(a) is a LEED pattern of the 4~u.c. thin film at room temperature, whereas Fig.~\ref{LEED}(b) shows a LEED pattern of 4~u.c. thin film at 16~K. Both LEED patterns are taken with the electron beam energy of 142~eV. White dotted rectangles represent primitive cells of a cubic SRO. The 4~u.c. SRO thin film shows $\sqrt{2}\times\sqrt{2}$ spots. In contrast, 50~u.c. SRO thin film shows $2 \times 2$ spots, indicating that the in-plane real space unit cell is enlarged two times with respect to that of the 4~u.c. SRO thin film.

RHEED patterns were also obtained after the growth of SRO thin film. Figures~\ref{LEED}(d) and (e) show RHEED patterns of 4 and 50~u.c. SRO thin film at room temperature when the electron beam direction is along the [100]$_c$ azimuth. Bragg spots at 0$^{th}$ and 1$^{st}$ Laue circle and Kikuchi lines are clearly observed, implying that our SRO thin films have high crystalline order at the surface. In Fig.~\ref{LEED}(d), only (-100), (000) and (100) Bragg peaks are observed whereas in Fig.~\ref{LEED}(e), (${1\over2}$00) and (-${1\over2}$00) Bragg peaks are observed additionally, also representing that the in-plane unit cell of 50~u.c. SRO thin film is doubled with respect to the 4 u.c. SRO thin film. A previous study showed that thick SRO films grown on the STO substrate at room temperature or lower have the orthorhombic structure ($a^{-}a^{-}c^{+}$), while SRO thin films thinner than 17~u.c. are stabilized with the tetragonal structure ($a^{0}a^{0}c^{-}$)~\cite{SHChang}. Our 4~u.c. SRO thin film also maintains the tetragonal structure ($a^{0}a^{0}c^{-}$) without any octahedral tilting at room temperature and 15~K, which is consistent with the COBRA experiment data.

\begin{center}
	{\bf Appendix D: Tilt angle-dependent Hall effect measurement on 4 u.c. SRO thin film}
\end{center}

\begin{figure}[htbp]
	\centering
	\includegraphics[width=0.48\textwidth]{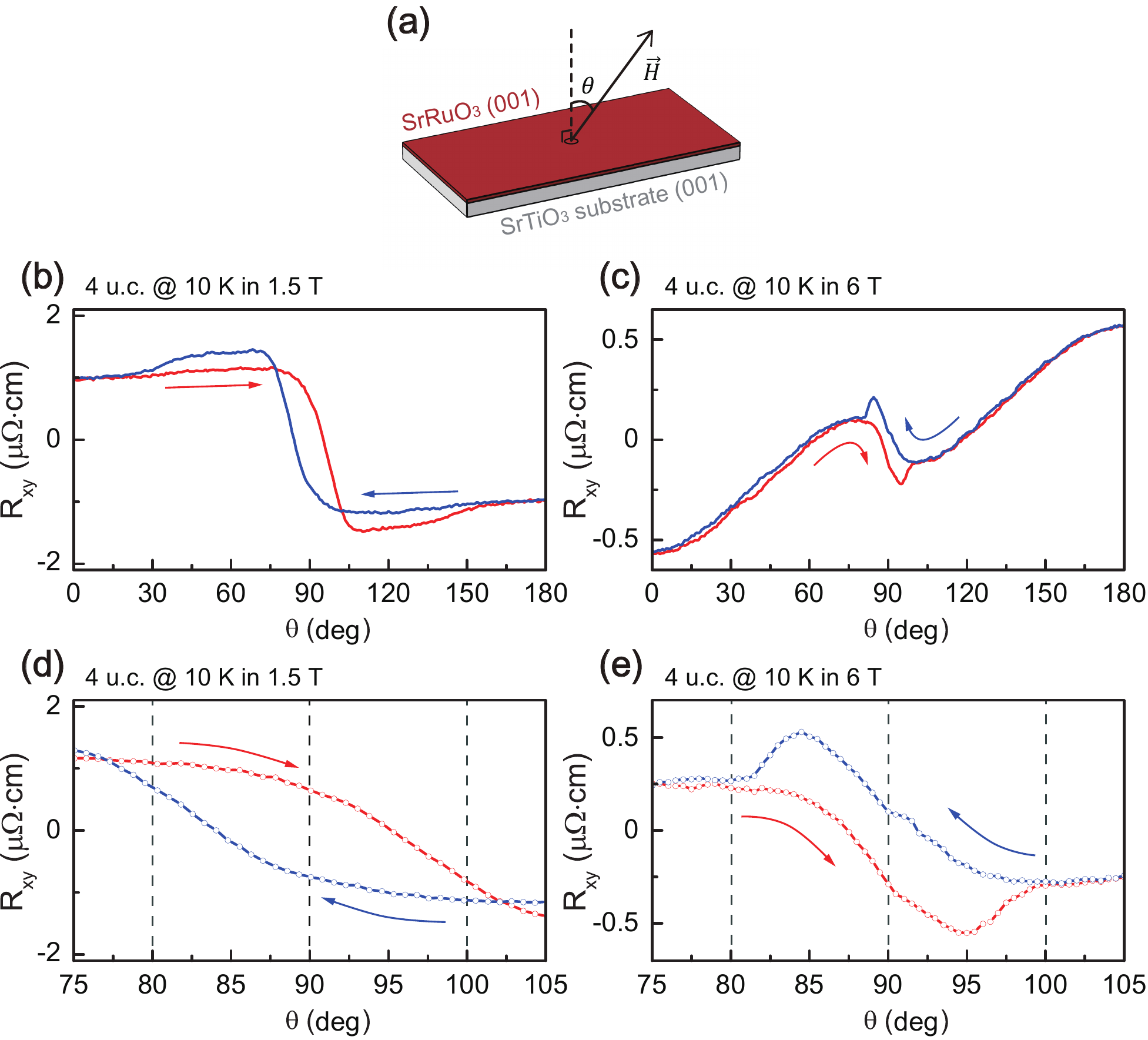}
	\caption{{\bf Angle-dependent Hall effect measurement on 4 u.c. SRO thin film.} (a) A schematic diagram of the angle-dependent Hall effect measurement. The angle between the external magnetic field and the normal direction of SRO thin film is $\theta$, and {\bf H} is the external magnetic field. (b), (c) Angle-dependent Hall resistance at 10 K with 1.5 T and 6 T external magnetic field. The red (blue) arrow indicates the positive (negative) sweep direction of the external magnetic field. (d), (e) An enlarged plot of (b) and (c). Black dotted lines are drawn to guide eyes to the fact that the hysteresis pattern is centered at 90$^\circ$. }
	\label{Angle}
\end{figure}

Figure~\ref{Angle}(a) shows the schematic setup for the angle-dependent Hall effect measurement. With the fixed external magnetic field ${\bf H}$, the 4~u.c. SRO is rotated and the Hall resistance is measured. The measured range of rotation is from 0$^{\circ}$ to 370$^{\circ}$, although only the measurement from 0$^{\circ}$ to 180$^{\circ}$ is shown. Figures~\ref{Angle}(b) and (c) show angle-dependent Hall resistance of the 4~u.c. film at 10~K at the external magnetic field of 1.5~T and 6~T, respectively. Clear hysteresis is found in both fields. Figures~\ref{Angle}(d) and (e) are enlarged plots of Fig.~\ref{Angle}(b) and (c), respectively. In both figures, the reversal of the magnetic moment takes place only after the field angle has exceeded 90$^\circ$ from the normal, consistent with the easy-axis nature of the magnetic anisotropy in thin-film SRO~\cite{easy-axis-1, easy-axis-2}.

\begin{center}
	{\bf Appendix E: Layer-dependent lattice elongation along $c$-axis and enhanced octahedral rotation in 4~u.c. SRO thin film observed by COBRA}
\end{center}

\begin{figure*}[htbp]
	\centering
	\includegraphics[width=0.9\textwidth]{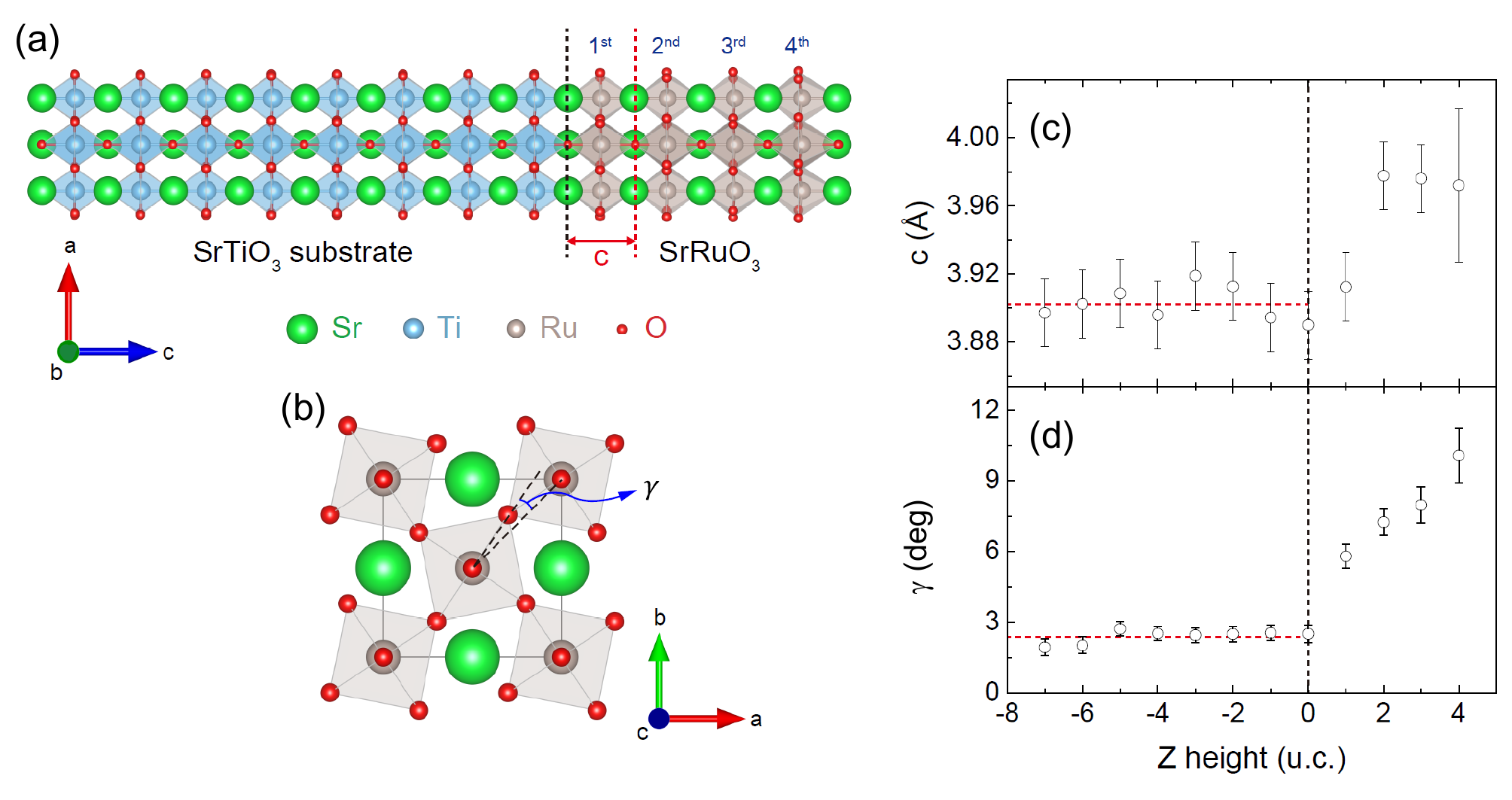}
	\caption{{\bf Layer-dependent c-axis lattice constant and octahedral rotation angle in 4 u.c. SRO thin film.} (a) A crystal model of SRO on the STO substrate. A green (blue, gray, red) dot indicates Sr (Ti, Ru, O) atom. A blue (gray) translucent octahedron indicates an oxygen octahedron surrounding the Ti (Ru) atom. SRO-STO interface is marked by black dotted line. c-axis lattice constant is defined as $c$. (b) \textit{ab}-plane atomic structure of SRO with the octahedral rotation angle $\gamma$ between the Ru-Ru and the Ru-O bond. (c) Layer-dependent $c$-axis lattice constant $c$ of the STO substrate and 4 u.c. SRO thin film. (d) Layer-dependent octahedral rotation angle $\gamma$ of the STO substrate and 4 u.c. SRO thin film.}
	\label{COBRA}
\end{figure*}

Layer-by-layer $c$-axis lattice constants and \textit{ab}-plane octahedral rotation angles in ultrathin SRO film can be obtained by COBRA combined with atomically resolved surface X-ray diffraction measurements. Figure~\ref{COBRA}(a) shows the atomic structure of the STO substrate and the 4~u.c. SRO thin film. The $c$-axis lattice constant, denoted $c$, is defined as the length between two adjacent Sr-O planes. The \textit{ab}-plane atomic structure of SRO thin film is given in Fig.~\ref{COBRA}(b). The octahedral rotation angle $\gamma$ is defined as the angle between the Ru-Ru bond and Ru-O bond. The $\gamma$ value is quantitatively calculated using the full width of the oxygen electron density peaks in the folded structure of the COBRA image~\cite{fister14}. Figures~\ref{COBRA}(c) and (d) show layer-by-layer lattice constants and octahedral rotation angles of STO substrate and the 4~u.c. SRO thin film. The average value of $c$ and $\gamma$ of the STO substrate measured at 30~K is 3.903~\AA~ and 2.401~\AA, respectively, as represented in the figures with red dotted lines. The experimental results clearly demonstrate that STO has a tetragonal phase at 30~K~\cite{loetz}.

Due to the compressive strain of the STO substrate, the \textit{ab}-plane lattice is suppressed and the lattice constant tends to be fixed to that of the STO substrate. As a result, SRO layers prefer to elongate toward the surface, causing the $c$-axis lattice constant to become larger than in the conventional pseudocubic SRO bulk. The oxygen octahedral rotation on the topmost layer of 4~u.c., at $\sim10^\circ$, is substantially larger than the 7.4$^\circ$  rotation angle found in the conventional SRO thin film grown on STO substrate at room temperature~\cite{gao16}. The COBRA result is quite consistent with our previous \textit{in-situ} LEED and RHEED studies and provides further detailed information.

\begin{figure}[htbp]
	\centering
	\includegraphics[width=0.5\textwidth]{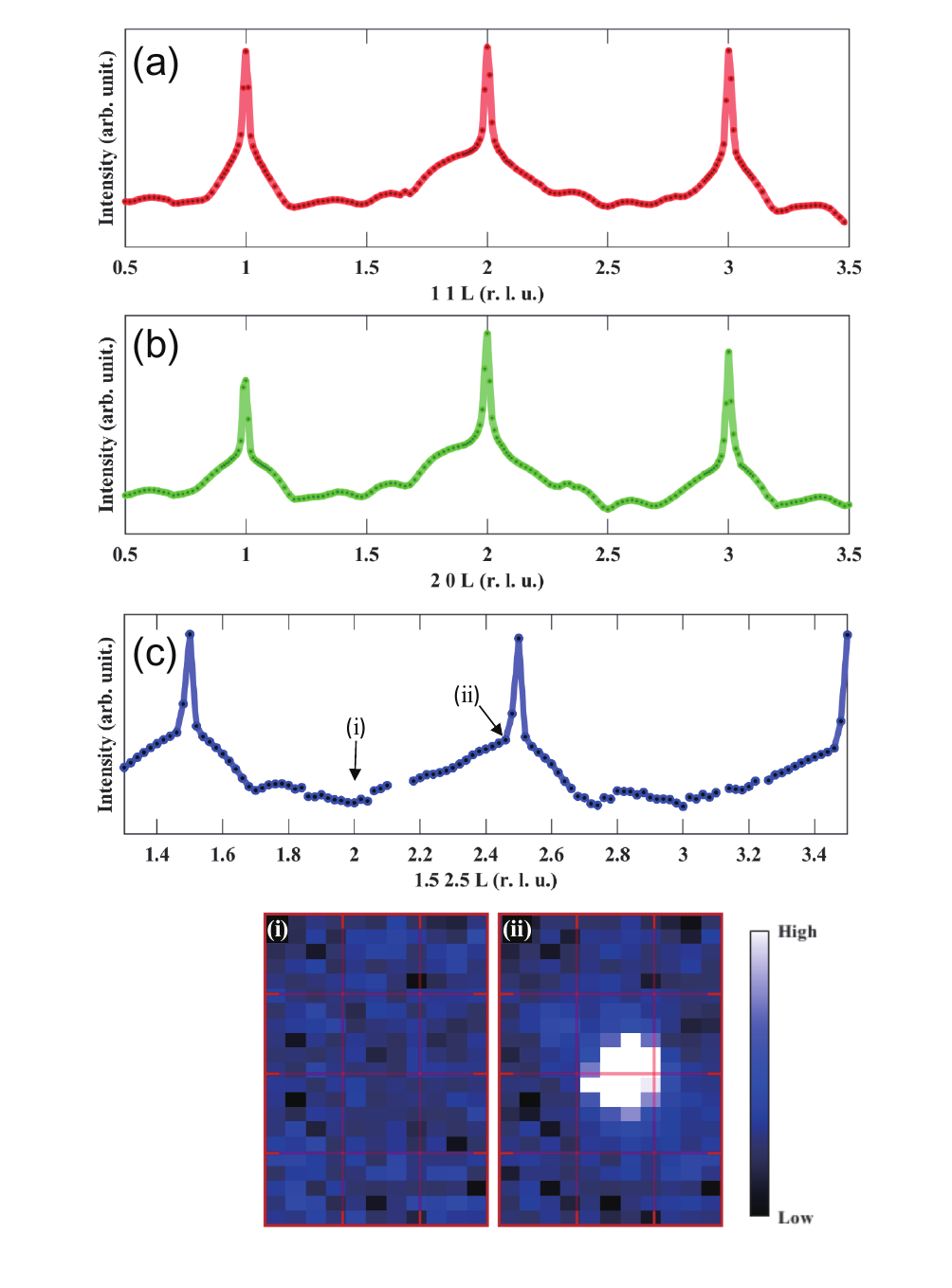}
	\caption{{\bf Representative off-specular crystal truncation rods for the COBRA } (a) $1$ $1$ $L$, (b) $2$ $0$ $L$, and (c) $1.5$ $2.5$ $L$ measured at 30 K. The symmetry of the crystal structure was consistent with the Half-order Bragg peak diffraction pattern. Examples of parts of two-dimensional detector images were taken at the (i) $L=2$ and (ii) $L=2.46$.}
	\label{COBRA2}
\end{figure}

In Fig.~\ref{COBRA2}, the crystal truncation rod (CTR) measurements for COBRA were carried out at 30~K at which the STO substrate has a tetragonal symmetry. The electron density profiles obtained by the COBRA method were indeed ‘folded’ electron density distributions coherently contributed by multiple structural domains. Using the folded electron density profile, we can extract the oxygen octahedral rotations ~\cite{fister14, yuan18}.

\begin{center}
	{\bf Appendix E: Calculating of the DMI energy using first-principles method}
\end{center}

In order to calculate the magnetic anisotropy of the film, we used (SrTiO$_{3}$)$_{4}$/(SrRuO$_{3}$)$_{4}$-vacuum configuration with experimental atomic positions in which an additional SrO layer was added to have SrO terminated surfaces for both ends and a vacuum of 15~\AA~ was used.  The magnetic anisotropy energy was calculated with spin-orbit coupling by comparing the total energy per Ru ion calculated with the magnetic moments fixed along the $c$-axis with that along the $a$-axis. The DMI vector in the $ab$-plane is obtained  by calculating the energy difference of four different spin configurations, mapped into a classical Hamiltonian with nearest neighbor spin interactions. The total energy of each spin configuration is calculated with $2\sqrt{2}\times 2\sqrt{2} \times 2$ supercell having the average octahedral distortion and rumpling of SRO layers from COBRA measurements of 4 u.c. SRO film in which the direction of the local spin moments of Ru ions is constrained by adding a penalty functional implemented in the VASP. The weigh in the penalty functional is increased until the contribution from the penalty function becomes less than 0.15~meV per supercell, less than 10~\% of the estimated DMI vector. The $4\times 4\times 6$ $k$-points grid are used for supercell with the increased energy cutoff of 700~eV.

We calculate the DMI interaction parameters following Ref. \cite{Xiang2011}. We consider the nearest-neighbor spin Hamiltonian explicitly written with DMI parameters:
\begin{eqnarray}
	H_{sp} = \sum_{\langle ij \rangle} \mathbf{D}_{ij} \cdot \mathbf{S}_{i}\times \mathbf{S}_{j} + H_{other}[\{ \mathbf{S}_{i} \}] ~,
\end{eqnarray}
where $\mathbf{S}_{i}$ denotes the spin moment at $i^{th}$ Ru site, $\mathbf{D}_{ij}$ is the DM vector, the bracket in the summation represents the nearest neighbors, and $H_{other}[\{ \mathbf{S}_{i} \}]$ denotes other spin interactions such as single ion anisotropy and exchange interactions. We define $z$-axis as the direction perpendicular to the interface and $x$- and $y$-axis along the in-plane pseudo-cubic axes. Due to the rumpling along the $z$-axis, we expect  the DM vectors connecting two Ru sites to be parallel to either $x$- or $y$-axis, depending on the direction connecting the two sites. We consider two Ru sites with site indices 1 and 2 connected along the $x$-direction having the non-zero DM vector in the $y$ direction, with which the spin Hamiltonian can be written as
\begin{eqnarray}
	\begin{split}
	H_{sp} = {D}^{y}_{12} S^{z}_{1} S^{x}_{2} - {D}^{y}_{12} S^{x}_{1} S^{z}_{2} +  \sum_{i \neq 2}\mathbf{D}_{1i} \cdot \mathbf{S}_{1}\times \mathbf{ S}_{i} + \\  \sum_{i\neq 1 }\mathbf{D}_{2i} \cdot \mathbf{S}_{2}\times \mathbf{ S}_{i} + H^{DM}_{other}[\{ \mathbf{S}_{i} \}] + H_{other}[\{ \mathbf{S}_{i} \}] ~,
	\end{split}
\end{eqnarray}
where $H^{DM}_{other}[\{ \mathbf{S}_{i} \}]$ represents the DM interaction between Ru spins excluding the sites 1 and 2.  We consider the total energy of the four spin configurations with local spin moment $S$: $(1)\; \mathbf{S}_{1} = (S, 0, 0), \mathbf{S}_{2} = (0, 0, S)$, $(2)\; \mathbf{S}_{1} = (S, 0, 0), \mathbf{S}_{2} = (0, 0, -S)$, $(3)\; \mathbf{S}_{1} = (-S, 0, 0), \mathbf{S}_{2} = (0, 0, S)$, and $(4)\; \mathbf{S}_{1} = (-S, 0, 0), \mathbf{S}_{2} = (0, 0, -S)$, while all other spin moments are aligned in the $y$ direction ($\mathbf{S}_{i} = (0,S,0)$ for $i \neq$ 1 or 2). We define the total energy of the $i^{th}$ spin configuration as $E^{i}$ and it can be easily shown that the value of the DM vector connecting the site 1 and 2 can be obtained by evaluating the energy difference \cite{Xiang2011}
\begin{eqnarray}
	D^{y}_{12} = \frac{1}{4S^{2}} (E^{1} +E^{4} - E^{2} -E^{3} )~.
\end{eqnarray}

The total energy of each spin configuration is calculated by constraining the magnetic moments of each Ru atom (see Methods in the main text) with a $2\sqrt{2}\times 2\sqrt{2} \times 2$ supercell having the average octahedral rotation and rumpling of 4 u.c. SRO film measured with COBRA. The obtained total energy relative to the smallest values are (8, 0, 22, 29) meV per supercell with average magnetic moment of Ru of 1.34~$\mu_{B}$, which gives the value of $D^{y}_{12}$ as 2.1~meV.

\end{document}